\documentclass[aps,10pt,prb,twocolumn, superscriptaddress]{revtex4-1}
\usepackage{graphicx}
\usepackage{amssymb, amsmath}
\usepackage{amsfonts}
\usepackage{commath}
\usepackage{color}
\usepackage{epstopdf}
\usepackage{multirow}
\usepackage{rotating}
\usepackage{footnote}
\usepackage{placeins}

\newcommand\varpm{\mathbin{\vcenter{\hbox{%
  \oalign{\hfil$\scriptstyle+$\hfil\cr
          \noalign{\kern-.3ex}
          $\scriptscriptstyle({-})$\cr}%
}}}}
\newcommand\varmp{\mathbin{\vcenter{\hbox{%
  \oalign{$\scriptstyle({+})$\cr
          \noalign{\kern-.3ex}
          \hfil$\scriptscriptstyle-$\hfil\cr}%
}}}}
\DeclareMathAlphabet      {\mathbf}{OT1}{cmr}{bx}{n}

\begin{document}

\title{Magnetic properties and magnetic structure of the frustrated quasi-one-dimensional antiferromagnet SrCuTe$_2$O$_6$}

\author{P.~Saeaun}
\affiliation{Department of Physics, Faculty of Science, Mahidol University, Bangkok 10400, Thailand}

\author{Y.~Zhao}
\affiliation{Department of Materials Science and Engineering, University of Maryland, College Park, Maryland 20742, USA}
\affiliation{NIST Center for Neutron Research, National Institute of Standards and Technology, Gaithersburg, Maryland 20899, USA}

\author{P.~Piyawongwatthana}
\affiliation{IMRAM, Tohoku University, Sendai, Miyagi 980-8577, Japan}

\author{T.~J.~Sato}
\affiliation{IMRAM, Tohoku University, Sendai, Miyagi 980-8577, Japan}

\author{F.~C.~Chou}
\affiliation{Center for Condensed Matter Sciences, National Taiwan University, Taipei 10617, Taiwan}
\affiliation{National Synchrotron Radiation Research Center, Hsinchu 30076, Taiwan}
\affiliation{Taiwan Consortium of Emergent Crystalline Materials, Ministry of Science and Technology, Taipei 10622, Taiwan}
\affiliation{Center of Atomic Initiative for New Materials, National Taiwan University, Taipei 10617, Taiwan}

\author{M.~Avdeev}
\affiliation{Australian Nuclear Science and Technology Organisation, ANSTO, Locked Bag 2001, Kirrawee DC, NSW, Australia}

\author{G.~Gitgeatpong}
\affiliation{Department of Physics, Faculty of Science and Technology, Phranakhon Rajabhat University, Bangkok 10220, Thailand}
\affiliation{Thailand Center of Excellence in Physics, Ministry of Higher Education, Science, Research and Innovation, 328 Si Ayutthaya Road, Bangkok 10400, Thailand}

\author{K.~Matan}
\email[Corresponding author: ]{kittiwit.mat@mahidol.ac.th}
\affiliation{Department of Physics, Faculty of Science, Mahidol University, Bangkok 10400, Thailand}
\affiliation{Thailand Center of Excellence in Physics, Ministry of Higher Education, Science, Research and Innovation, 328 Si Ayutthaya Road, Bangkok 10400, Thailand}

\date{\today}
\begin{abstract}
Magnetization measurements on single-crystal cubic SrCuTe$_2$O$_6$ with an applied magnetic field of along three inequivalent high symmetry directions $[100]$, $[110]$, and $[111]$ reveal weak magnetic anisotropy. The fits of the magnetic susceptibility to the result from a quantum Monte Carlo simulation on the Heisenberg spin-chain model, where the chain is formed via the dominant third-nearest-neighbor exchange interaction $J_3$, yield the intra-chain interaction $(J_3/k_B)$ between 50.12(7)~K for the applied field along $[110]$ and 52.5(2)~K along $[100]$ with about the same $g$-factor of 2.2.  Single-crystal neutron diffraction unveils the transition to the magnetic ordered state as evidenced by the onset of the magnetic Bragg intensity at $T_\textrm{N1}=5.25(9)$~K with no anomaly of the second transition at $T_\textrm{N2}$ reported previously.  Based on irreducible representation theory and magnetic space group analysis of powder and single-crystal neutron diffraction data, the magnetic structure in the Shubnikov space group $P4_132$, where the Cu$^{2+}$~$S=1/2$ spins antiferromagnetically align in the direction perpendicular to the spin chain, is proposed. The measured ordered moment of $0.52(6)~\mu_B$, which represents 48\% reduction from the expected value of $1~\mu_B$, suggests the remaining influence of frustration resulting from the $J_1$ and $J_2$ bonds.
\end{abstract}
\maketitle

\section{Introduction}

During the past few decades, quantum magnetism in low-dimensional and frustrated systems have captured the interest of condensed-matter physicists because of their potential to exhibit exotic magnetic ground states such as spin ice~\cite{Castelnovo2008, Kadowaki:2009, Fennell415, Bramwell2009}, a quantum valence bond (dimer) solid~\cite{PhysRevLett.59.799,Kageyama1999,Matan2010}, and the most sought-after quantum spin-liquid~\cite{Balents2010, Savary:2017fk, RevModPhys.88.041002}.
Among these unconventional states, a quantum spin liquid has gained the most attention because its discovery and fundamental understanding could potentially yield a better understanding of other phenomena in physics such as high-$T_c$ superconductivity~\cite{Anderson:1973eo, *Anderson:1987ii}, topological states, and anyonic physics~\cite{Kitaev2006}, and lead to applications in quantum computing~\cite{RevModPhys.80.1083}.  

A search for this elusive quantum spin liquid has so far focused on low-dimensional and frustrated lattices, which include triangular-based lattices.  Topologically, triangular (two-dimensional (2D) edge-sharing-triangle), kagome (2D corner-sharing-triangle), hyper-kagome (3D corner-sharing triangle), and pyrochlore (3D edge-sharing-tetrahedron) lattices are considered as possible hosts of the quantum-spin-liquid ground state due to their high degree of frustration, giving rise to a macroscopically degenerate ground state, which could prompt the formation of a highly entangled quantum state. In recent years, significant progress has been made in the search, and many possible realizations of the quantum spin liquid were discovered and exhaustively studied~\cite{Broholmeaay0668}. A group of triangular-based materials that show quantum-spin-liquid traits includes ZnCu$_3$(OH)$_6$Cl$_2$~\cite{Shores:2005de,Helton2007,Han2012}, Na$_4$Ir$_3$O$_8$~\cite{Okamoto2007}, YbMgGaO$_4$~\cite{Li:2015bt, Paddison:2017jm}, Ca$_{10}$Cr$_{7}$O$_{28}$~\cite{RN16417, PhysRevB.97.104413}, Ce$_2$Zr$_2$O$_7$~\cite{RN3}, and PbCuTe$_2$O$_6$ (Cu$^{2+}, S=1/2)$~\cite{Kote2014,Khuntia2016}. 

Studies of PbCuTe$_2$O$_6$ revealed the absence of a long-range $\lambda$-like transition down to 2 K and possible emergence of quantum spin liquid at low temperatures~\cite{Kote2014,Khuntia2016,Chillal2017}.  On the other hand, SrCuTe$_2$O$_6$ with an almost identical crystal structure exhibits two successive magnetic phase transitions, one of which is to a magnetically ordered state at the N\'eel temperature $T_\textrm{N1}$ of 5.5 K~\cite{Ahmed2015,Kote2015}. In order to decipher the underlying mechanism that gives rise to different ground states in these two seemingly similar systems, a detailed study on a single-crystal sample is required. Hence, in this article we report the studies of magnetic properties and a magnetic structure on powder and single-crystal SrCuTe$_2$O$_6$.

\begin{figure*}
\includegraphics[width=0.85\textwidth,trim={0mm 0mm 0mm 0mm},clip]{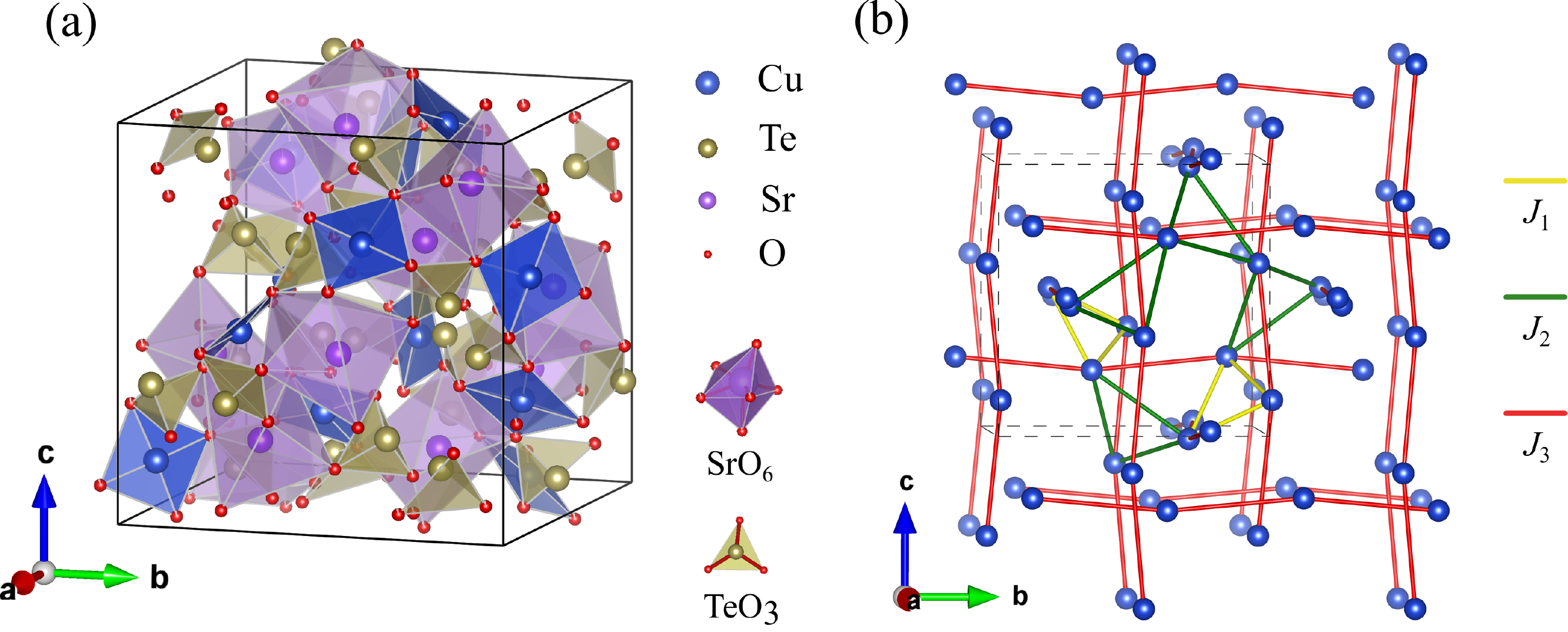}
\caption{\label{fig1}(a) The crystal structure of SrCuTe$_2$O$_6$ consists of CuO$_4$ square plaquettes (blue), TeO$_3$ triangular plaquettes (yellow), and SrO$_6$ octahedra (purple). (b) The intertwined spin network of Cu$^{2+}$~$S=1/2$ spins is connected by the exchange interactions $J_1$ (yellow),  $J_2$ (green), and  $J_3$ (red) to form isolated triangles, a hyper-kagome lattice, and spin chains, respectively.}
\end{figure*}

Noncentrosymmetric SrCuTe$_2$O$_6$ crystallizes in the chiral cubic space group $P4_132$ (No.~213) with lattice parameter $a$ = $12.473(1)$ \AA~\cite{Wulff1997}. The crystal structure consists of CuO$_4$ square plates, units of TeO$_3$, and oxygen octahedra surrounding strontium (Sr) atoms as shown in Fig.~\ref{fig1}(a). The Cu$^{2+}$ ions in the CuO$_4$ plaquettes carry spin 1/2 and give rise to the magnetic properties of this system. As with PbCuTe$_2$O$_6$, the spin network of SrCuTe$_2$O$_6$ consists of three types of intertwined networks connected by the nearest-, second-nearest-, and third-nearest-neighbor exchange interactions $J_1$, $J_2$, and $J_3$, respectively~\cite{Ahmed2015,Kote2015}, as shown in Fig.~\ref{fig1}(b). The Cu$^{2+}$ ions connected by $J_1$ form an isolated triangle with Cu-O-Sr-O-Cu superexchange pathways. The second-nearest-neighbor network formed by $J_2$ via Cu-O-Te-O-Cu pathways connect the Cu$^{2+}$ ions to form a hyper-kagome network. Finally, $J_3$ links the Cu$^{2+}$ ions to form spin chains along the cubic crystallographic axes.  First principle calculations performed to estimate the strength of these exchange interactions showed that for SrCuTe$_2$O$_6$, $J_3$ is the most dominant with $J_2$ about 10\% of $J_3$ and  $J_1$ about 1\% of $J_3$, and established that SrCuTe$_2$O$_6$ is a spin-chain system with relatively weak and intricate inter-chain couplings~\cite{Ahmed2015}.  In contrast, for PbCuTe$_2$O$_6$, $J_2$ is the most dominant interaction, and hence this system can be characterized as the 3D hyper-kagome antiferromagnet, in which geometric frustration can potentially suppress the N\'{e}el state and prompt the emergence of the quantum spin state~\cite{Kote2014,Khuntia2016,Chillal2017}.

Previous magnetic susceptibility and heat capacity measurements on a SrCuTe$_2$O$_6$ powder sample~\cite{Ahmed2015,Kote2015,Kote2016} revealed magnetic transitions at $T_\text{N1}=5.5$ K, which is characterized as the paramagnetic-to-antiferromagnetic transition, and an unexplained anomaly at $T_\text{N2}$, which is smaller than $T_\textrm{N1}$ and field-dependent~\cite{Ahmed2015,Kote2015}. The magnetic susceptibility data also showed a broad maximum, which is typical in a spin-chain system, at around 32 K suggesting short-range correlation.  A fit of the high-temperature data to the Curie-Weiss law yields the Curie-Weiss temperature $\Theta_\text{CW} = -35$ K.  Assuming that the spin-chain network is connected by $J_3$, the dominant exchange interaction ($J_3/k_B$) is estimated to be 49~K as compared to the first-principle-calculation value of 45~K~\cite{Ahmed2015}.  Furthermore, multiple magnetic transitions with a nontrivial $H-T$ phase digram were observed for magnetizations up to 9 T~\cite{Ahmed2015,Kote2015}.

In this work, magnetic properties of SrCuTe$_2$O$_6$ are studied using magnetization measurements on a single-crystal sample, and the magnetic transition at $T_\textrm{N1}$ and magnetic structure of the ordered state are investigated using neutron diffraction. This article is organized as follows.  In Section~\ref{sec1}, the single-crystal synthesis and experimental methods  are described. The result of the single crystal X-ray diffraction is discussed in~\ref{sec2a}, macroscopic magnetic properties on the single crystals are investigated and analyzed using quantum Monte Carlo simulations in~\ref{sec2b}, and powder and single-crystal neutron diffraction data are discussed in ~\ref{sec2c}.  The article ends with the conclusion in Section~\ref{sec3}

\section{Experimental details}\label{sec1}

A powder sample of SrCuTe$_2$O$_6$ was first synthesized by the standard solid state reaction of high purity SrCO$_3$, TeO$_2$ and CuO. The preparation method is described elsewhere~\cite{Kote2015}. The obtained pure phase powder of SrCuTe$_2$O$_6$ was then used as a starting material for single crystal growth using the vertical gradient freeze technique. The powder was loaded into a pointed-bottom alumina tube (recrystallized alumina 99.8\%). The crucible was then sealed in an evacuated quartz tube, which is crucial to minimize the formation of SrCuTe$_2$O$_7$ as an impurity phase. The sample was melted at 800 $^\circ$C and held at this temperature for 24 hours to ensure a homogeneous and complete melt, before subjecting the sample to a 20 $^\circ$C/cm temperature gradient in a vertical furnace at a rate of 1 cm/day. After the sample reached the position with $T=$ 650 $^\circ$C, the furnace was cooled to room temperature at a rate of 200 $^\circ$C/h. The sample was then mechanically extracted from the crucible.  Single crystals with the largest mass of about 1 g were obtained.

Small pieces of the crystals were collected and ground thoroughly for a powder X-ray diffraction measurement to confirm sample purity. The powder X-ray diffraction data was fitted using the Rietveld method implemented in \textsc{fullprof}~\cite{Carvajal1993}.  Single-crystal X-ray diffraction was also performed using a Bruker X8 APEX II CCD Diffractometer with Mo$K\alpha$ radiation at room temperature. The refinement on the single crystal diffraction data for fractional coordinates was done using \textsc{ShelXle}~\cite{Hubschle:kk5092}.

In order to investigate the macroscopic magnetic properties of SrCuTe$_2$O$_6$ in the single crystal sample, the magnetic susceptibility was measured with the applied field aligned along the three inequivalent directions $[100]$, $[110]$, and $[111]$. A single crystal was aligned using a four-circle X-ray diffractometer and cut into a cube with dimensions of 1$\times$1$\times$1 mm$^3$ ($\sim$20 mg).  The aligned crystal was then attached to a Teflon rod using GE-7031 varnish and placed inside a measuring stick.  The magnetic susceptibility was measured as a function of temperature from 2~K up to 300~K with the applied magnetic field of 1.0~T using a superconducting quantum interference device (MPMS-XL, Quantum Design). To analyze the susceptibility data, quantum Monte Carlo (QMC) simulations were performed to capture the broad maximum of the magnetic susceptibility data in order to extract the value of the leading exchange interaction. The QMC simulations were performed using the \textsc{loop} algorithm~\cite{Todo2001} in the \textsc{alps} simulation package~\cite{Bauer_2011} on a cluster of 100 spins for the spin-chain model and 96~000 spins ($20\times20\times20$ unit cells) for the $J_3$ model of SrCuTe$_2$O$_6$ in the temperature range of $0.01 \leqslant t \leqslant 10$ $(t = k_\textrm{B}T/J)$ with 100~000 Monte Carlo steps for thermalization and 500~000 Monte Carlo steps after thermalization.  The numerical result was fitted with the experimental data by a Pad{\'{e}} approximant~\cite{Johnson2000} and the leading exchange interaction and Land\'e $g$-factor were finally obtained.

To study the magnetic structure, powder neutron diffraction was performed at the high-resolution neutron powder diffractometer BT1 at the NIST Center for Neutron Research (NCNR), USA.  The Ge(311) monochromator was used to select neutrons with $\lambda = 2.077$~\AA~and collimations of $60'-20'-7'$ were employed. In addition, elastic neutron scattering on a small 130 mg single crystal was conducted using the Double Focusing Triple-Axis Spectrometer BT7~\cite{BT7} at NCNR. Neutron scattering measurement was performed using fixed incident energy of 14.7 meV and pyrolytic graphite PG(002) was used as a monochromator. The sample was cooled to the base temperature using a closed cycle $^4$He cryostat. A position sensitive detector (PSD) was employed in the two-axis mode with collimations of $\textrm{open} - 80' - 80' - \textrm{PSD}$. The sample was aligned so that the $hhl$ plane was in the scattering plane.

\section{Results and discussion}\label{sec2}
\subsection{X-ray diffraction}\label{sec2a}

\begin{figure*}
 \includegraphics[width=1.0\textwidth,trim={0mm 0mm 0mm 0mm},clip]{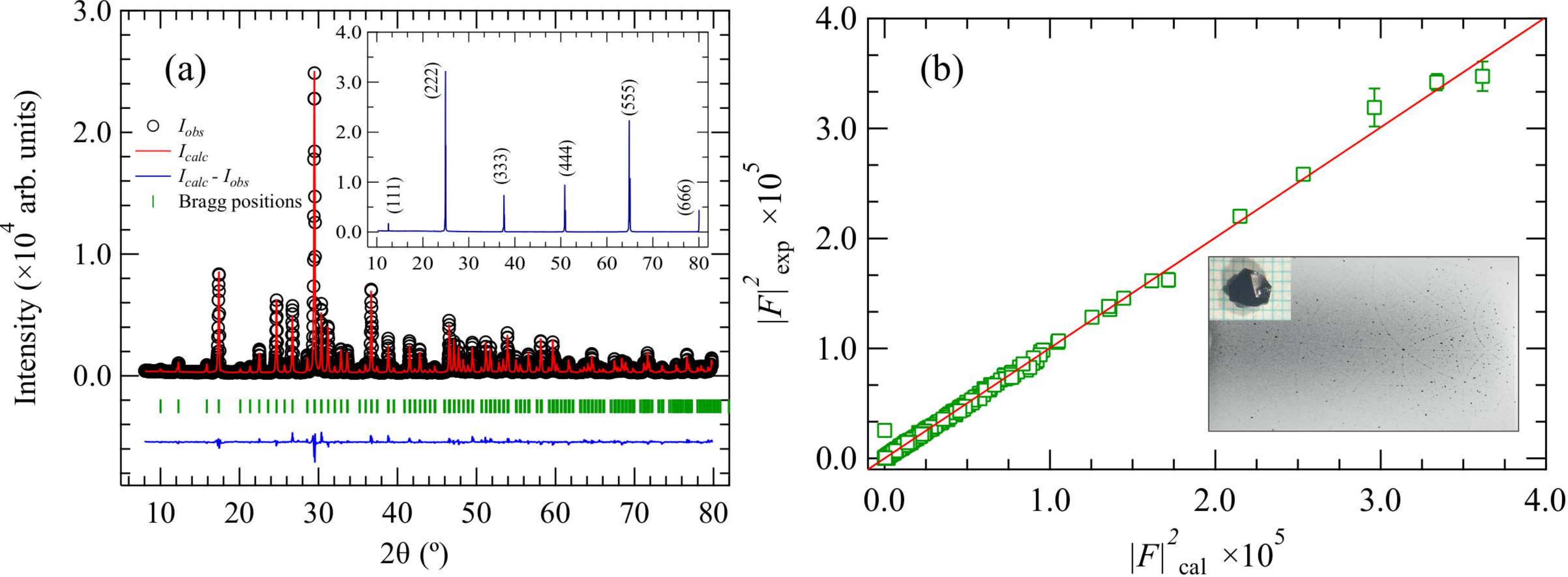}
 \caption{\label{fig2}(a) Powder X-ray diffraction pattern measured on the crushed single crystals of SrCuTe$_2$O$_6$ at room temperature. The black circle denotes the observed data, the red line the calculated pattern, the vertical green line the Bragg positions, and the blue line the difference between the observed and calculated patterns. The inset shows the $\theta$-2$\theta$ scan on the cleaved  $(1,1,1)$ facet of single crystal SrCuTe$_2$O$_6$.  (b) The refinement result of the single crystal X-ray diffraction data shows the agreement between the measured and calculated scattering intensities. Error bars represent one standard deviation. The inset shows a photograph of a crystal and Laue X-ray diffraction measured on the cleaved surface.}
 \end{figure*}

The X-ray diffraction data, which were measured on the powder sample obtained by grinding small pieces of the single crystals, along with the result of the Rietveld refinement are shown in Fig.~\ref{fig2}(a). The results show that the obtained SrCuTe$_2$O$_6$ single crystals are single-phase without any trace of impurity.  However, we will later show that from the neutron diffraction data measured on an as-grown powder sample, SrCuTe$_2$O$_7$ is present as an impurity. The absence of this impurity phase in the single crystal X-ray diffraction is due to the fact that only small pieces of the SrCuTe$_2$O$_6$ single crystals were selected for grinding, and hence SrCuTe$_2$O$_7$, which also showed up as having a different color, was selected out. The lattice parameter obtained from the refinement is $a=12.466(6) $ \AA\ for the cubic space group $P4_132$ in agreement with previous reports~\cite{Wulff1997,Ahmed2015,Kote2015,Kote2016}. The residual parameters for the Rietveld refinement are $R_p=6.66\%$, $R_{wp}=8.49\%$, and the goodness of fit (GoF) of 2.1.  A $\theta$-2$\theta$ scan shown in the inset of Fig.~\ref{fig2}(a) on the naturally cleaved facet confirms single-crystallinity and shows that the cleaved facet is the $111$ plane. To further investigate the quality of the single crystals, Laue photography was also performed in the transmission mode. The result shown in the inset of Fig.~\ref{fig2}(b) reveals clear Bragg peaks confirming the quality of the crystal. 

Single crystal X-ray diffraction at room temperature was also performed and the data were refined to extract the atomic fractional coordinates. The refinement was done on 1192 unique reflections with $F_{obs}>4\sigma(F_{obs})$ yielding the fitted lattice parameter $a = 12.472(6)$~\AA. Figure~\ref{fig2}(b) shows the agreement between the measured and calculated intensity resulting from the refinement with the residual parameters $R_1=2.47\%$, $wR_2=6.01\%$, and GoF=1.042.  We note that $R_1$ for the refinement performed on all (1263) reflections is 2.79\%. The fractional coordinates of SrCuTe$_2$O$_6$ obtained from the refinements of the powder (the crushed crystalline sample) and single-crystal X-ray diffraction data are summarized in Appendix~\ref{Appendix1}.
 
\subsection{Magnetic susceptibility}\label{sec2b}

\begin{figure*}
 \includegraphics[width=1.0\textwidth,trim={0mm 0mm 0mm 0mm},clip]{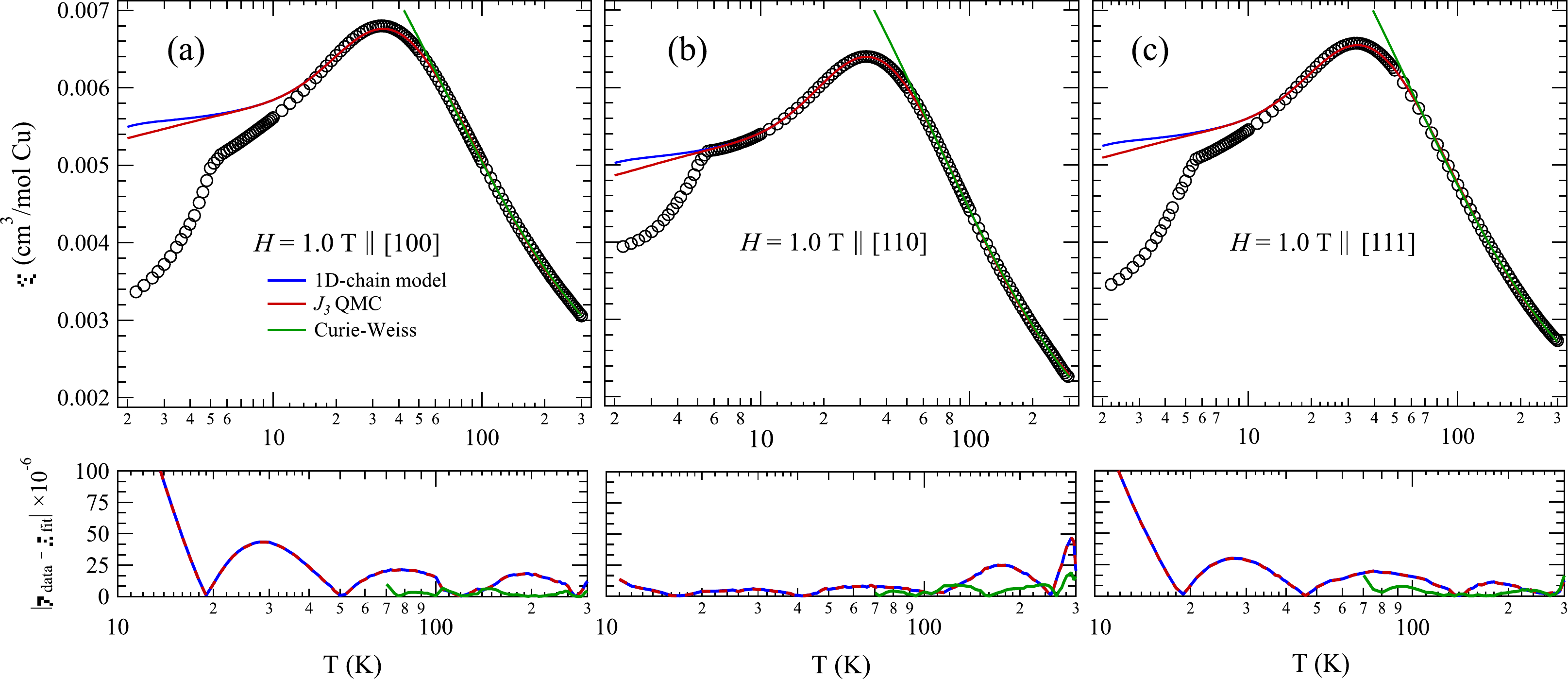}
 \caption{\label{fig3} Magnetic susceptibility measured on a single crystal of SrCuTe$_2$O$_6$ with applied magnetic fields of 1.0 T along three crystallographic directions, (a) $[100]$, (b) $[110]$ and (c) $[111]$. The blue and red lines represent the fits to the 1D-chain model and our QMC simulations, respectively.  The green line denotes the Curie-Weiss-law fit.  The lower panels show the residuals between the data and the fits.}
 \end{figure*}

\begin{figure}
 \includegraphics[width=0.48\textwidth,trim={0mm 0mm 0mm 0mm},clip]{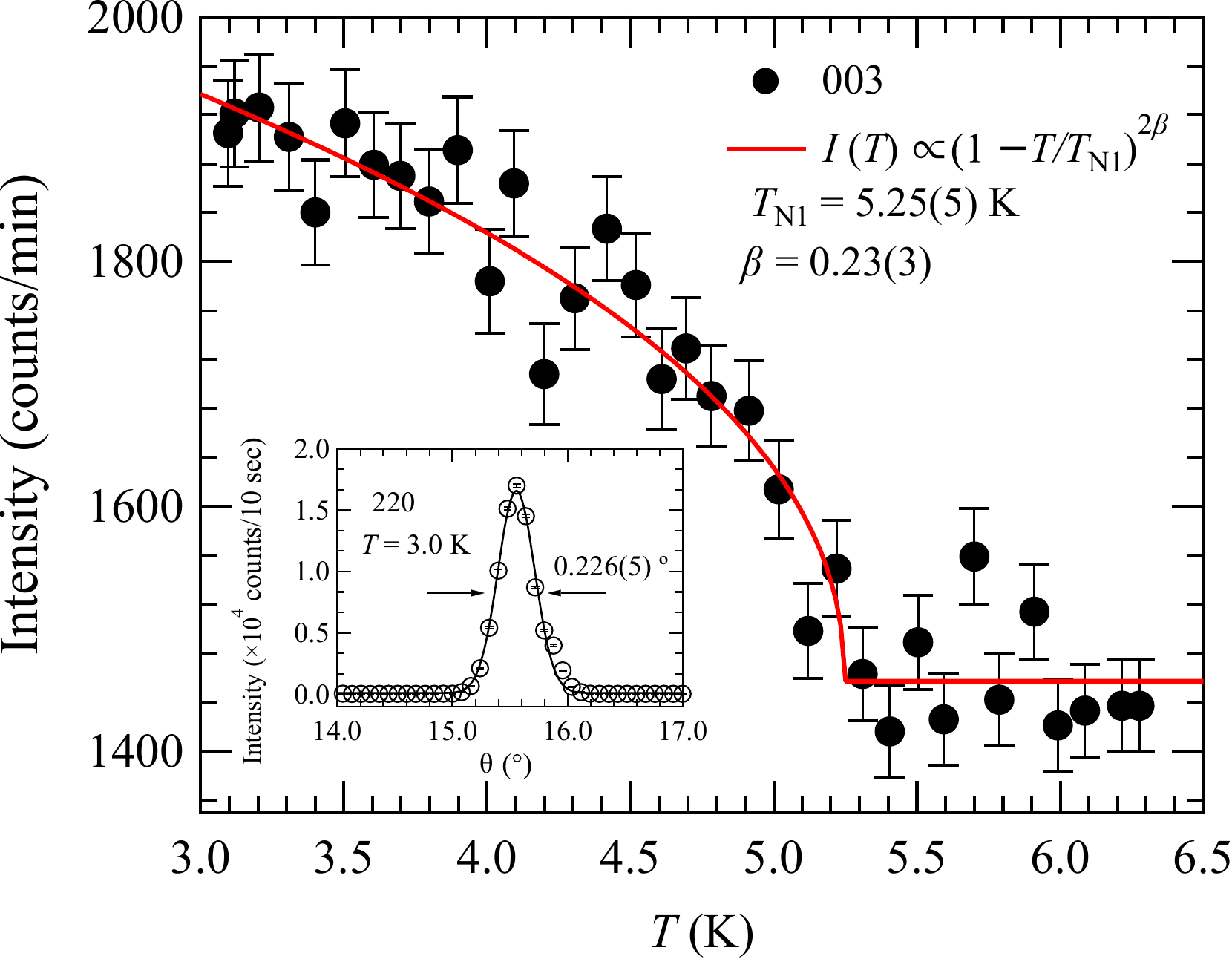}
 \caption{\label{fig4} Neutron diffraction intensity of the $003$ reflection as a function of temperature. The red line denotes a power-law fit for the magnetic scattering intensity representing the order parameter. Error bars represent one standard deviation. The inset shows the $\theta$-scan of the $220$ structural reflection.  We note that error bars are smaller than the plot symbol. The solid line represents the Gaussian fit to capture the line-shape of the peak. }
 \end{figure}

Figure~\ref{fig3} shows the magnetic susceptibility measured with the applied field along three inequivalent directions $[100]$, $[110]$, and $[111]$.  The susceptibility data are plotted against the temperature in a log scale to emphasize the broad maximum at $\sim$30~K, and the kink at the N{\'e}el temperature $T_{\text N1}$ = 5.25(5) K ($T_\textrm{N1}$ was obtained from the neutron scattering data shown in Fig.~\ref{fig4}). Below $T_\textrm{N1}$, the magnetic susceptibility decreases, which is indicative of the antiferromagnetic arrangement of the spins. The broad maximum is a sign of short-range spin correlation, typical for low-dimensional antiferromagnetic systems~\cite{PhysRev.135.A640}, and can be used to estimate the leading exchange interactions of the system. This broad peak was consistent with that previously observed in the powder sample~\cite{Ahmed2015,Kote2015}. We note that for the data measured in the applied field of 1.0 T on the powder sample in Ref.~\onlinecite{Ahmed2015,Kote2015}, the anomaly at $T_\text{N2}$ is absent, possibly because it is very weak at low field; for the magnetization measurements, this anomaly, which is field-dependent, appears below $T_\textrm{N1}$ for $H\geq2.0$~T~\cite{Ahmed2015}. Consistent with the previous work on the powder sample, we were able to observe the second magnetic transition at higher magnetic fields (not shown here).  

 \begin{table}
 \caption{\label{table2}The fitted parameters of magnetic susceptibilities data along $[100]$, $[110]$, and $[111]$ with the mean-field Curie-Weiss law (Eq. \ref{CWlaw})}
 \begin{tabular*}{0.48\textwidth}{@{\extracolsep{\fill}}cccc}
 \hline
 \hline
$H  \parallel$ & $\chi_0~(\textrm{cm}^3/\text{mol Cu}$)  (fixed) & $\Theta_{\text{CW}}$ (K) & $\mu_\textrm{eff}$~($\mu_\textrm{B}$) \\  
\hline
$[100]$ & $1.69\times10^{-3}$ &     $-46.5(6)$   &    1.97(1) \\
$[110]$ & $0.834\times10^{-3}$ &     $-41.6(8)$   &    2.01(2)  \\
$[111]$ & $1.31\times10^{-3}$  &     $-44.3(6)$   &    1.99(1) \\
\hline    
\hline
 \end{tabular*}
 \end{table}
 
\begin{table}
 \caption{\label{table3}The fitted parameters of the magnetic susceptibilities data along $[100]$, $[110]$, and $[111]$ with the QMC simulation}
 \begin{tabular*}{0.48\textwidth}{@{\extracolsep{\fill}}cccc}
\hline
\hline
$H \parallel $ &  $\chi_0~(\text{cm}^3/\text{mol Cu})$ & $J/k_B$ (K) & $g\text{-factor}$\\ 
\hline  
$[100]$ & $1.69(1)\times10^{-3}$  &   $52.5(2)$  &   2.197(6)\\
$[110]$ & $0.834(5)\times10^{-3}$ &  $50.12(7)$ &  2.249(2) \\
$[111]$ & $1.31(1)\times10^{-3}$  &   $51.8(1)$  &   2.221(5)\\ 
\hline
\hline
 \end{tabular*}
 \end{table}

To estimate the dominant exchange coupling between the Cu$^{2+}$~$S=1/2$ spins and verify the connectivity of the spin network, we fit the high-temperature $(T\gtrsim T_\textrm{N1})$ data to the mean-field and QMC calculations. The magnetic susceptibility data above 70 K along the three axes were first fitted with the mean-field Curie-Weiss law,
\begin{equation}
\chi(T) = \chi_0 + \frac{C}{T-\Theta_\text{CW}}, \label{CWlaw}
\end{equation}
where $C$ and $\Theta_\text{CW}$ are the Curie constant and Curie-Weiss temperature, respectively.  $\chi_0$, which is fixed to the values obtained from the fits of the QMC results (later discussed), represents the temperature-independent background, which is relatively high (about half of the measured susceptibility at high temperatures) due to the Teflon used to fix the crystal.  We note that $\chi_0$ is positive and its magnitude is about an order of magnitude larger than the core diamagnetic and Van Vleck paramagnetic susceptibilities~\cite{Ahmed2015}. Teflon is diamagnetic and hence gives rise to negative magnetic susceptibility.  However, in the measurements, the signal from the Teflon rods was treated as background by MPMS.  As a result, the empty gap between the Teflon rods gives rise to positive magnetic-susceptibility background.  The difference in the value of $\chi_0$ could be due to the difference in the gap between the two Teflon rods that were used to fix the aligned crystal in different orientations. The Curie-Weiss fitted curves are shown by the green lines in Figs.~\ref{fig3}(a)-(c). The fitted parameters along each field direction are shown in Table~\ref{table2}. The effective magnetic moment, $\mu_\textrm{eff} = \sqrt{3k_\textrm{B}C/N_\textrm{A}}$, obtained for the three inequivalent directions does not significantly deviate from one another, which is indicative of relatively weak spin anisotropy with a similar $g-$factor.  All are found to be slightly higher but still close to the spin-only value of $\mu_\textrm{eff}  = g\mu_\textrm{B}\sqrt{S(S+1)} = 1.73~\mu_\textrm{B}$ for $g = 2$ and $S = 1/2$. The slight increase of the value of $\mu_\textrm{eff}$ from the spin-only value could indicate the presence of the spin-orbit coupling and the orbital contribution to the magnetic moment.  For $S=1/2$, the measured $\mu_\textrm{eff}$ of 2.0 implies $g=2.3$. The obtained Curie-Weiss temperature, which is close to the previous reported values measured on the powder sample~\cite{Ahmed2015,Kote2015}, as shown in Table \ref{table2}, indicates no significant deviation along the three field directions. The negative Curie-Weiss temperature indicates that the dominant exchange interaction is antiferromagnetic. The order of frustration $f = |\Theta_\textrm{CW} / T_\textrm{N1}|$ for $T_\textrm{N1} = 5.25$ K and the average Curie-Weiss temperature of $-44$~K, yields $f \sim 8$ suggesting some degree of frustration in this system that could result from $J_1$ and $J_2$ that give rise to the frustrated spin networks, isolated triangles and the hyper-kagome lattice, respectively. 

The decrease of the magnetic susceptibility for $T<T_\textrm{N1}$ suggests an antiferromagnetically ordered state with the absence of ferromagnetism, which in some materials could be due to spin canting.  The absence of the spin canting is consistent with the absence of the staggered Dzyaloshinskii-Moriya (DM) interactions, which can give rise to canted moments, between two spins connected by $J_3$.  However, the chiral $P4_132$ symmetry of the underlying crystal structure can lead to the uniform DM interactions, which, in combination with the complex spin network, can result in the nontrivial $H-T$ phase diagram observed in previous work~\cite{Ahmed2015}.  It was discovered that a chiral material with space group $P4_132$ can host magnetic skyrmions~\cite{RN16418}, the existence of which can be explained by the DM interaction.  In addition, the similar drop of the magnetic susceptibility for $T\lesssim T_\textrm{N1}$ along all three field directions [Figs.~\ref{fig3}(a)-(c)] further suggests that there is no global easy axis and the antiferromagnetic alignment must be along a local easy axis determined by the local environment around the Cu$^{2+}$ ions. 

To go beyond the mean-field approximation, we performed a QMC simulation to calculate the magnetic susceptibility that was subsequently used to fit the experimental data. The previous work showed that the broad maximum observed in SrCuTe$_2$O$_6$ can be well captured by the 1D spin-chain model~\cite{Ahmed2015,Kote2015}. For this work, we performed QMC simulations with the \textsc{loop} algorithm based on the one-dimensional spin-chain model and on the spin network connected by $J_3$ [shown in Fig.~\ref{fig1}(b)].  We note that the two models without any inter-chain coupling are in fact equivalent.  However, the $J_3$ spin network was used for possible inclusion of $J_1$ and $J_2$, which serve as inter-chain interactions.  The simulated result was then fitted to the experimental data using
\begin{equation}
\chi(T)=\chi_0+\chi_\textrm{QMC}(T), 
\end{equation}
where
\begin{equation}
\chi_\textrm{QMC}(T) = \frac{N_\textrm{A}\mu_\textrm{B}^2g^2}{k_\textrm{B}J}\chi^\ast (k_\textrm{B}T/J).
\end{equation}
$N_\textrm{A}$, $\mu_\textrm{B}$, and $k_\textrm{B}$ are the Avogadro constant, Bohr magneton, and Boltzmann constant, respectively.  $J$ and $g$ are fitted to the experimental data.  The susceptibility $\chi^\ast(t)$ as a function of the reduced temperature $t=k_\textrm{B}T/J$ was obtained by fitting the QMC results using a Pad\'e approximant~\cite{Johnson2000, cvo1}.  The fitting was performed on the experimental data from $T =$ 300 K down to 10 K, which is slightly above $T_\textrm{N1}$ and contains the broad peak, along the three field directions as shown in Fig.~\ref{fig3}.  The QMC result appears to fit the data for the applied field along the $[110]$ direction very well, where the QMC calculations and the experimental data are in good agreement around the broad peak and the consistency extends down to the temperature just above $T_\textrm{N1}$.  In contrast, for the $[100]$ and $[111]$ data, the QMC results appear to deviate from the data around the broad peak.  We cannot explain why the result of the QMC calculations fits the $[110]$ data much better than it does the other two field directions. The fitted parameters are summarized in the Table \ref{table3}. The fitted values of the Land{\'e} $g$-factor are consistent with those obtained from the Curie-Weiss fit discussed in the previous section. It can be seen that the obtained exchange interactions along three axes do not deviate much and the values are very close to those obtained from the mean-field approximation. This suggests that the 1D spin-chain model along third-nearest neighbors adequately describes the macroscopic magnetic properties. The obtained Land{\'e} $g$-factor along the three inequivalent field directions from the QMC fit are also consistent with the spin network where the chains are running along the three crystallographic axes giving rise to the relatively isotropic magnetic susceptibility in this cubic system.  We attempted to include $J_1$ and $J_2$ in our QMC calculations but encountered the sign problem due to the frustration of the $J_1$ and $J_2$ bonds.  As a result, we were unable to obtain reliable $J_1$ and $J_2$ values especially around the broad peak and the low-temperature region.  

\begin{figure*}
 \includegraphics[width=1.0\textwidth,trim={0mm 0mm 0mm 0mm},clip]{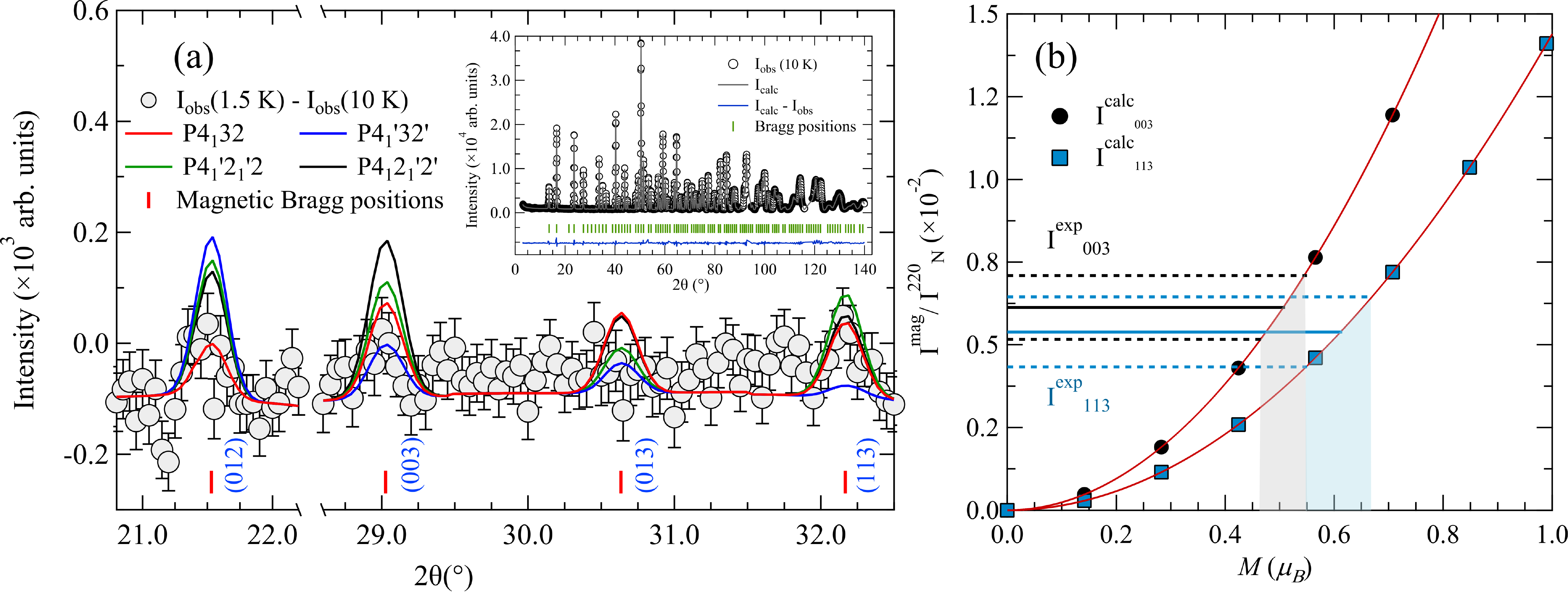}
 \caption{\label{fig5}(a) The difference data between 1.5 K and 10 K show the magnetic Bragg scattering at $012$, $003$, $013$, and $113$. The lines represent the magnetic-structure refinements based on magnetic space groups $P4_1'32'$, $P4_132$, $P4_12_1'2'$, and $P4_1'2_1'2$.  The vertical red lines denote the magnetic Bragg reflections. The powder neutron diffraction data were measured at 10 K (the inset).  The black symbols denote the observed data, the grey line the calculated nuclear-scattering pattern, the vertical green lines the Bragg positions, and the blue line the difference between the observed and calculated data. (b) The calculated single-crystal magnetic intensities for the magnetic space group $P4_132$ are shown as a function of ordered moment $(M)$ for the reflections $003$ (black circles) and $113$ (blue squares). The red lines denote the fits to the quadratic function.  The black (blue) horizontal line represents the experimental value of the ratio between the magnetic scattering intensity measured at $003$ and $113$, and the structural scattering intensity measured at $220$. The gray (blue) shaded region indicates a range of the measured ordered moment for $003$ and $113$.}
 \end{figure*}

\subsection{Single-crystal and powder neutron diffraction}\label{sec2c}

To investigate the microscopically magnetic properties of SrCuTe$_2$O$_6$, elastic neutron scattering was performed on a small crystal ($\approx$ 130 mg) at BT7. The inset of Fig.~\ref{fig4} shows a $\theta$-scan (rotating only the sample) around the $220$ structural Bragg peak with a Gaussian fit yielding a full-width-at-half-maximum (FWHM) equal to 0.226(5)$^\circ$, indicative of good crystallinity. The scattering intensity of the $003$ and $113$ reflections  was measured as a function of temperature.  The results show the onset of the intensity increase at $T_\textrm{N1}$.  The extra scattering intensity below $T_\textrm{N1}$ is indicative of magnetic scattering resulting from the ordering of the magnetic moments.  Figure~\ref{fig4} shows the $003$ scattering intensity data, which also serve as a measure of the order parameter below $T_\textrm{N1}$, along with the power law fit. The fit to $I(T)\propto(1-T/T_\textrm{N1})^{2\beta}$ for the order parameter yields the critical exponent $\beta = 0.23(3)$ and Ne\'{e}l temperature $T_{\text N1} = 5.25(5)$~K, where the errors represent one standard deviation.  We note that the magnetic scattering intensity is proportional to $M^2$, where $M$ is the sublattice magnetic moment, and hence the factor of two in the exponent. For the $113$ data (not shown), the critical exponent  and N\'{e}el temperature are 0.27(4) and 5.12(5) K, respectively, which are consistent with those obtained from the $003$ data. The fitted value of $\beta$ is typical for low-dimensional magnetic systems~\cite{Tennant1995,Kojima1997,Banks2009}, the 1D spin chain in this case. The obtained value of $T_{\text N1}$ is consistent with magnetic susceptibility data and with previous reports on the powder sample~\cite{Ahmed2015,Kote2015,Kote2016}. However, given the resolution of the data, the order parameter in Fig.~\ref{fig4}, which was measured at zero magnetic field, does not show an anomaly of the second magnetic transition around $T_\text{N2}$. It is possible that the second transition can only be detectable at high field where the anomaly becomes stronger as suggested by the magnetization data.  Hence in order to investigate the magnetic structure change at $T_\text{N2}$, future in-field elastic neutron scattering is required.

To determine the magnetic structure of SrCuTe$_2$O$_6$, powder neutron diffraction was performed at BT1 and the data were collected at 10 K and 1.5 K, above and below $T_\textrm{N1}$, respectively. The refinement of the nuclear structure was first performed on the 10-K data with the fitted lattice parameter $a=12.4321(5)$~\AA.  The resulting Rietveld refinement [Fig.~\ref{fig5}(a)], which yields $R_\text{p}$~=~3.72\%, shows results that are consistent with the crystal structure of SrCuTe$_2$O$_6$ obtained from the X-ray diffraction data (Appendix~\ref{Appendix1}). The sample, however, contained some impurities, the majority of which was identified to be SrCuTe$_2$O$_7$ that constitutes roughly 2.1 wt.\%. The proximity of impurity reflections to some of the magnetic Bragg reflections and weak magnetic intensity hinder the refinement of the magnetic structure from the powder neutron diffraction data.  As a result, the fitted ordered magnetic moment has large error as will be discussed below.

\begin{figure*}
 \includegraphics[width=0.8\textwidth,trim={0mm 0mm 0mm 0mm},clip]{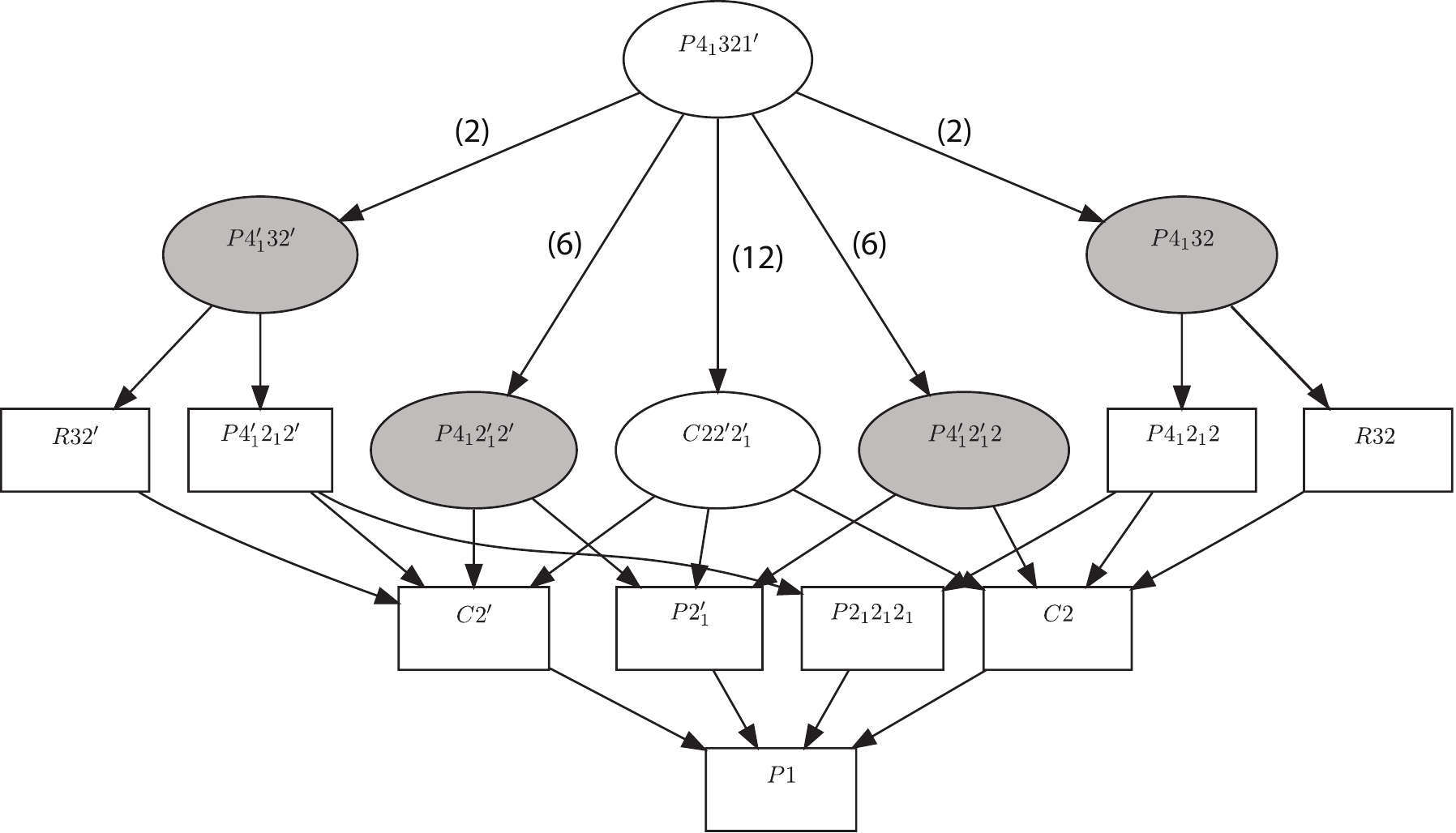}
 \caption{\label{fig6} A diagram of subgroups shows a hierarchy of possible subgroups of the paramagnetic parent space group $P4_1321'$.  The maximal subgroups are indicated by eclipses, and those that were used in the refinement are highlighted by shading. A subgroup index between the parent space group and a maximal subgroup is shown in a parenthesis. The diagram is generated using $k$-\textsc{subgroupsmag}~\cite{Perez-Mato2015}.}
 \end{figure*}

The magnetic structure of SrCuTe$_2$O$_6$ was first analyzed using the irreducible representation theory. The detail was described in Appendix~\ref{Appendix2}. We note that due to the large number of free parameters, we were unable to perform full refinement for $\Gamma_3$, $\Gamma_4$, and $\Gamma_5$ with 6, 12 and 15 free parameters, respectively.  Therefore, we have to rely on the magnetic space group analysis in order to further sub-classify possible magnetic structures of SrCuTe$_2$O$_6$ and reduce the number of free parameters.

Based on the Landau-type transition with a single order parameter, the magnetic Shubnikov space groups can be derived from the paramagnetic parent space-group $P4_1321'$ giving rise to 14 Shubnikov space groups as shown by the graph of subgroups, which was generated using $k$-\textsc{subgroupsmag}~\cite{Perez-Mato2015}, in Fig.~\ref{fig6}. Out of these 14 subgroups, there are a total of five maximal magnetic subgroups, $P4_132$ (No. 213.63), $P4_1'32'$ (No. 213.65), $P4_12_1'2'$~(No. 92.114), $P4_1'2_1'2$~(No. 92.115), and $C22'2_1’$~(No. 20.34). $P4_132$ and $P4_1'32'$ correspond to $\Gamma_1$ and $\Gamma_2$, respectively, whereas $P4_12_1'2'$, $P4_1'2_1'2$, and $C22'2_1'$ correspond to $\Gamma_4$ and $\Gamma_5$.  Since some of the basis vectors of $\Gamma_4$ and $\Gamma_5$ are absent for the magnetic space groups $P4_12_1'2'$, $P4_1'2_1'2$, and $C22'2_1'$, the number of fitting parameters is reduced.

Assuming that symmetry reduction at the magnetic transition to the ordered state is minimal, we performed the refinement of the magnetic structure on the 1.5-K data for the four (out of five) maximal subgroups, namely $P4_132$ ($\Gamma_1$) with one free parameter, $P4_1'32'$ ($\Gamma_2$) with 2 free parameters, $P4_12_1'2'$ with 5 free parameters, and $P4_1'2_1'2$ with 4 free parameters.  We note that an attempt to perform the refinement for the maximal subgroup $C22'2_1'$ with 9 free parameters was unsuccessful.  All fitting parameters including the lattice parameter, atomic positions, and peak profile parameters, were kept constant and the same as those obtained from the fitting of the 10-K data.  However, the background was adjusted due to the difference in the incident neutron flux. Since the magnetic scattering was observed at low momentum transfer, and hence low $2\theta$ angles, the magnetic-structure refinement was performed for $20.5^\circ< 2\theta <22.5^\circ$ and $28.5^\circ< 2\theta <32.5^\circ$, where four magnetic Bragg reflections were observed. The inset of Fig.~\ref{fig5}(a) shows the difference pattern between 1.5~K and 10~K for these four magnetic reflections, $012$, $003$, $013$ and $113$, two of which [$003$ and $013$] appear next to the impurity peaks (not shown). The figure also shows that the calculated intensity based on the Shubnikov space group $P4_132$, which yields the R-factor of 5.51\%, fits the data slightly better than that based on $P4_1'32'$ with the R-factor of 7.02\%.  The fitted magnetic moment for $P4_132$ is 0.8(7) $\mu_\textrm{B}$, where the error was estimated from fitting the 10-K data; the large error is due to the significant contribution from the nearby impurity peaks and weakness of the magnetic signal. The refinement based on $P4_12_1'2'$ and $P4_1'2_1'2$ yields the R-factors of 6.31\% and 6.38\%, respectively, which are marginally worse than that for $P4_132$. We note that even though the number of fitting parameters for $P4_1'32'$, $P4_12_1'2'$, and $P4_1'2_1'2$ is higher than that for $P4_132$, the fitted results for $P4_1'32'$, $P4_12_1'2'$, and $P4_1'2_1'2$ do not become better.  However, the small difference of the R-factors is not statistically significant enough to validate that SrCuTe$_2$O$_6$ magnetically orders in the magnetic space group $P4_132$.  Furthermore, we were unable to rule out $C22'2_1'$, nor, under the assumption of minimal symmetry reduction at the transition, examine non-maximal subgroups.  As previously noted, since the magnetic scattering in SrCuTe$_2$O$_6$ is weak and some magnetic Bragg reflections are in close proximity to impurity peaks, the refinement of the powder neutron diffraction data yields an inconclusive result with large error.  Hence, in this work, $P4_132$ is proposed as the most likely candidate based on the magnetic space group analysis.  

\begin{figure}
 \includegraphics[width=0.48\textwidth,trim={0mm 0mm 0mm 0mm},clip]{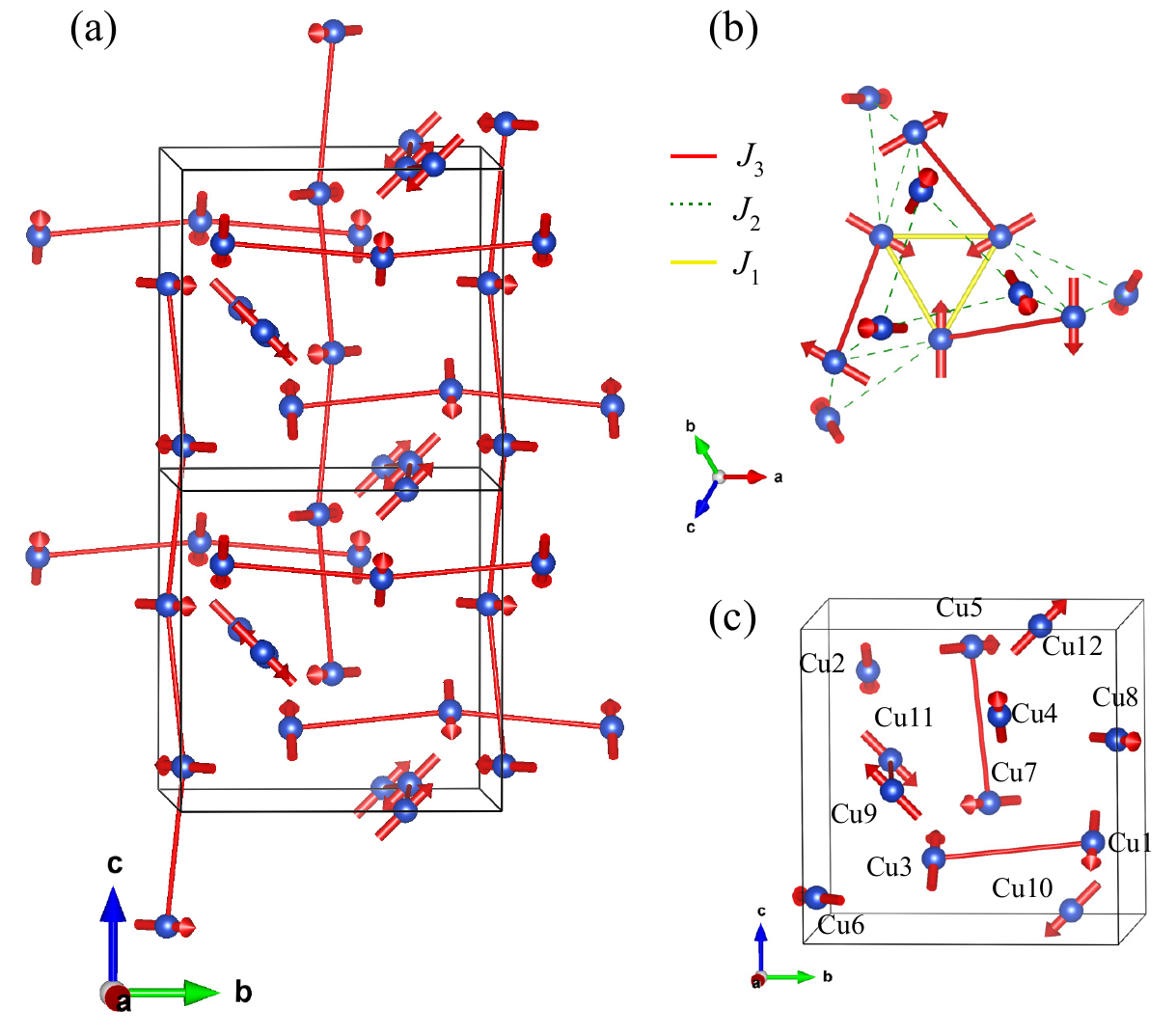}
 \caption{\label{fig7} (a) The magnetic structure of SrCuTe$_2$O$_6$ belongs to the Shubnikov space group $P4_132$ ($\Gamma_1$).  The dominant $J_3$ antiferromagnetic exchange interactions form spin chains along the crystallographic axes. (b) The spin network, which is formed by $J_1$ (yellow), $J_2$ (dashed green) and $J_3$ (blue), is shown along with the spin structure. (c) The Cu atoms in a unit cell are labeled according to Table~\ref{table4}.}
 \end{figure}

The resulting magnetic structure for $P4_132$ is shown in Fig.~\ref{fig7}. The magnetic moments of the Cu$^{2+}$~$S=1/2$ spins in SrCuTe$_2$O$_6$ antiferromagnetically align in the direction perpendicular to the chain formed by $J_3$, which is the most dominant exchange interaction, consistent with proposed spin network deduced from the DFT calculations~\cite{Ahmed2015}.  This antiferromagnetic spin structure is consistent with the magnetization data (discussed above) where no weak ferromagnetism, which could result from spin canting, was observed. Interestingly, we observed that the spins on the corners of an isolated triangle connected by $J_1$, {\it i.e.} the weakest exchange interaction among the three considered, form a co-planar 120$^\circ$ configuration [Fig.~\ref{fig7}(b)], which relieves, to some degree, the geometrical frustration inherent in the triangle-based spin network.   However, the $J_2$ interactions, which form the hyper-kagome spin network, remain highly frustrated.  The magnetic order in this quasi-1D system is most likely stabilized by this intricate network of further-nearest-neighbor interactions [Figs.~\ref{fig1}(b) and \ref{fig7}(b)].  

In order to better extract the value of the ordered moment, we performed the analysis on the single-crystal neutron diffraction data measured at BT7.  In Fig.~\ref{fig4}, we were able to clearly observe the magnetic Bragg intensities at $003$ and $113$ (not shown), which we will denote as $I_{M}^{003}$ and $I_{M}^{113}$, respectively. The base temperature of 3~K for the single-crystal experiment might not be low enough relative to $T_\textrm{N}$ to give a good estimate of the ordered moment as suggested by the increasing trend of the scattering intensity in Fig.~\ref{fig4}.  Hence, $I_{M}^{003}$ and $I_{M}^{113}$ were obtained by extrapolating the order parameter fitted curve of the scattering intensity measured at 003 and 113, respectively, to 1.5~K, at which the power neutron diffraction was measured. In comparison with the nuclear Bragg intensity at $220$, $I_{N}^{220}$, we calculated the ratio between the magnetic and nuclear scattering intensity as $I_{M}^{003}/I_{N}^{220} = 6.1(10)\times10^{-3}$ and $I_{M}^{113}/I_{N}^{220} = 5.4(11)\times10^{-3}$. These obtained values of the magnetic to nuclear intensity ratio are very weak {\it i.e.}, roughly of the same order of magnitude as the statistical error in the powder data, and hence small contributions from the nearby impurity scattering can cause a large error in the refinement.  As a result, the magnetic scattering is barely noticeable in the difference plot between the 1.5-K and 10-K data (Fig.~\ref{fig5}(a)), and the magnetic structure refinement on the powder neutron diffraction data fails to yield a reliable result. 

To obtain the value of the ordered magnetic moment from the single-crystal data, we compare the magnetic intensities $I_{M}^{003}/I_{N}^{220}$ and $I_{M}^{113}/I_{N}^{220}$ from the single crystal data and the magnetic scattering intensity calculated from \textsc{fullprof} for $P4_132$. We convert the integrated intensities calculated from \textsc{fullprof} for powder to those for single-crystal by multiplying $\sin\theta/m_{hkl}$, where $m_{hkl}$ is the multiplicity of the $hkl$ reflection~\cite{shirane2002}. The result is shown in Fig.~\ref{fig5}(b); the red curves in Fig.~\ref{fig5} denote a fit to $I^{\textrm{mag}}/I^{\textrm{220}}\propto M^2$.  We note that $M$ is the only free parameter for $P4_132$. Given the values of $I_{M}^{003}/I_{N}^{220} $ and $I_{M}^{113}/I_{N}^{220}$ from above, we estimate the ordered moment to be $0.51(4)\mu_\textrm{B}$ and $0.61(6)\mu_\textrm{B}$, respectively. The horizontal black and blue solid lines denote the values of the intensity ratios for $003$ and $113$, respectively, with the dashed lines representing the range of the error. We have done a similar analysis for $P4_1'32'$, which has two free parameters, $C_1$ and $C_2$.  We found that if $C_2$ is equal to zero, the calculated magnetic intensity of $113$ will be greater than that of $003$, which is inconsistent with the experimental data.  With increasing $C_2$, the $003$ intensity can become larger than the $113$ intensity but they are still inconsistent with the experimental data.  In addition, refinement of the single-crystal data was also performed using \textsc{Jana2006}~\cite{JANA2006} for magnetic space groups $P4_132$.  We were unable to check $P4_1'32'$, $P4_12_1'2'$, and $P4_1'2_1'2$ due to a limited number of data points. The least-square refinement yields $wR_2=4.63\%$ and goodness of fit (GoF) of 2.46.  The fitted ordered magnetic moment of $0.52(6)\mu_B$ is in good agreement with the above values obtained from the graph in Fig.~\ref{fig5}(b).  Hence, we reach the same conclusion as from the analysis of the powder data that the magnetic space group for the magnetically ordered state of SrCuTe$_2$O$_6$ below $T_\textrm{N1}$ is $P4_132$.  

The obtained ordered moment of $0.52(6)\mu_\textrm{B}$ is about half of the expected value of $1~\mu_\textrm{B}$ for $S=1/2$ ($48\%$ reduction), suggesting that frustration in the $J_1$ and $J_2$ bonds potentially induces spin fluctuations and significantly reduce the ordered moment. The reduction of ordered moments has been observed in ordered frustrated systems, KFe$_3$(OH)$_6$(SO$_2$)$_4$ $(S=3/2)$ with 24\% reduction~\cite{PhysRevB.61.12181}, and Cs$_2$Cu$_3$SnF$_{12}$~$(S=1/2)$ with 32\% reduction~\cite{PhysRevB.99.224404}.  In comparison, for PbCuTe$_2$O$_6$, where the dominant $J_2$ forms the frustrated hyper-kagome lattice, spin fluctuations are so large that the N\'eel state is totally suppressed and a quantum spin liquid possibly emerges at low temperatures~\cite{Kote2014,Khuntia2016,Chillal2017}.  From the DFT calculations, the exchange interactions in PbCuTe$_2$O$_6$ are close to one another, which could enhance the frustration whereas the intra-chain interaction $J_3$ in SrCuTe$_2$O$_6$ is, respectively, one order and two orders of magnitude larger than $J_1$ and $J_2$, which could place SrCuTe$_2$O$_6$ away from the quantum spin liquid state even though structurally it is almost identical to PbCuTe$_2$O$_6$.  Nevertheless, even though SrCuTe$_2$O$_6$ magnetically orders at low temperatures, the residue effect of the frustrated bonds remains and evidences in the reduced ordered moment.  It would be interesting to investigate this subtle effect of frustration in spin dynamics of this system.
 
\section{Conclusion}\label{sec3}

Magnetization measurements on single-crystal SrCuTe$_2$O$_6$ reveal highly isotropic magnetic susceptibility along the three inequivalent directions $[100]$, $[110]$, and $[111]$ in this cubic system. The value of the leading exchange interaction $(J_3/k_B)$ estimated using a quantum Monte Carlo simulation on the 1D spin-chain model is between 50.1 and 52.5 K.  The order parameter measured by neutron scattering confirms that the system magnetically orders below $T_\textrm{N1}=5.25(5)$~K. However, from our zero-field neutron diffraction mesurements, we are unable to confirm the field-dependent second transition at lower $T_{N2}$, which was previously observed in magnetization and heat capacity measurements. Further in-field neutron scattering measurements are required to investigate this second transition. Based on the neutron diffraction data on the powder and single-crystal samples, the magnetic structure in the Shubnikov space group $P4_132$, where the Cu$^{2+}$~$S=1/2$ spins antiferromagnetically align along the direction perpendicular to the spin chain, is proposed with the ordered magnetic moment of $0.52(6)\mu_\textrm{B}$. This work suggests the dominance of the intra-chain interaction $J_3$ over the frustrated $J_1$ and $J_2$ bonds, and sheds light on the difference in magnetic ground states between SrCuTe$_2$O$_6$ and PbCuTe$_2$O$_6$.  The 48\% reduction of the ordered moment in SrCuTe$_2$O$_6$ points to the residual effect of frustration, which could have nontrivial influence on spin dynamics in this magnetically ordered system.
\\[2mm]

{\it Note:} After submitting this manuscript, we became aware of similar work~\cite{chillal2020magnetic}, which was published in a public archive.  Magnetic susceptibility measured on a single-crystal sample reported in Ref.~\onlinecite{chillal2020magnetic} is consistent with our results.  Ref.~\onlinecite{chillal2020magnetic} also confirms our reported magnetic structure with the reduced ordered magnetic moment.  However, their measured value of the ordered moment is slightly lower than that reported in this work.
\begin{acknowledgements}
Work at Mahidol University was supported in part by the Thailand Research Fund (TRF) Grant Number RSA6180081 and the Thailand Center of Excellence in Physics.  PS was supported by the RGJ-PhD scholarship (Grant No. PHD/0114/2557) from TRF. FCC acknowledges funding support from the Ministry of Science and Technology (108-2622-8-002-016 and 108-2112-M-001-049-MY2) and the Ministry of Education  (AI-MAT 108L900903) in Taiwan.  The identification of any commercial product or trade name does not imply endorsement or recommendation by the National Institute of Standards and Technology.
\end{acknowledgements}

\bibliography{ref}
\clearpage
\appendix
\section{Atomic coordinates of SrCuTe$_2$O$_6$} \label{Appendix1}
The refined fractional atomic coordinates of SrCuTe$_2$O$_6$ were shown in Table \ref{table1}. The refinement of the powder (crushed crystalline sample) and single crystal data was performed using \textsc{fullprof}~\cite{Carvajal1993} and \textsc{ShelXle}~\cite{Hubschle:kk5092}, respectively. 

\begin{table}
\caption{\label{table1} Refined values of fractional coordinates of SrCuTe$_2$O$_6$ from powder and single crystal X-ray diffraction measured at room temperature, and powder neutron diffraction measured at 10 K. }
\centering
\begin{tabular*}{0.45\textwidth}{@{\extracolsep{\fill}}ccccc}
\hline
\hline
Atom & Site & $x/a$ & $y/a$ & $z/a$ \\
\hline
\multicolumn{5}{c}{powder X-ray diffraction}\\
Te & 24e &0.3380(1) & 0.9187(1)  & 0.0588(1) \\
Sr(1)  & 8c &0.0545(2)  & 0.0545(2)  & 0.0545(2)\\
Sr(2)  & 4b &0.375  &  0.625  &  0.125  \\
Cu  & 12d &0.4762(1)  & 0.875  &  0.2738(3)  \\
O(1) & 24e &0.6635(9) &  1.1271(9)  & 0.1761(9)  \\
O(2)  & 24e &0.4404(9)  & 1.0205(9)  & 1.2210(8)  \\
O(3)  & 24e &0.2222(9) &  0.9781(10)  & 0.1302(11)  \\
\multicolumn{5}{c}{{\sl R$_{p}$} = 0.0666, {\sl R$_{wp}$} = 0.0849, GoF = 2.1}\\
\hline
\multicolumn{5}{c}{single-crystal X-ray diffraction}\\
Te & 24e &0.33827(3) & 0.91872(3)  & 0.05938(3) \\
Sr(1)  & 8c &0.05469(5)  & 0.05469(5)  & 0.05469(5) \\
Sr(2)  & 4b &0.375  &  0.625  &  0.125  \\
Cu  & 12d &0.47567(7)  & 0.875  &  0.27433(7)  \\
O(1) & 24e &0.6710(4) &  1.1273(4)  & 0.1785(4)  \\
O(2)  & 24e &0.4382(4)  & 1.0171(4)  & 1.2284(4)  \\
O(3)  & 24e &0.2224(5) &  0.9767(6)  & 0.1305(5)  \\
\multicolumn{5}{c}{{\sl R$_{1}$} = 0.0247, {\sl wR$_{2}$} = 0.0601, GoF = 1.042}\\
\hline
\multicolumn{5}{c}{powder neutron diffraction}\\
Te & 24e &0.3379(1) & 0.9192(1)  & 0.0589(1) \\
Sr(1)  & 8c &0.0536(1)  & 0.0536(1)  & 0.0536(1) \\
Sr(2)  & 4b &0.375  &  0.625  &  0.125  \\
Cu  & 12d &0.4760(1)  & 0.875  &  0.2741(1)  \\
O(1) & 24e &0.6702(1) &  1.1271(1)  & 0.1795(1)  \\
O(2)  & 24e &0.4391(1)  & 1.0163(1)  &1.2271(1)  \\
O(3)  & 24e &0.2220(1) &  0.9766(2)  &0.1297(1)  \\
\multicolumn{5}{c}{{\sl R$_{p}$} = 0.0372 , {\sl R$_{wp}$} = 0.0487, GoF = 2.5}\\
\hline
\hline
\end{tabular*}
\end{table}

\section{Table of magnetic irreducible representations of SrCuTe$_2$O$_6$ } \label{Appendix2}

The magnetic structure of SrCuTe$_2$O$_6$ was analyzed using the irreducible representation theory. The analysis based on the symmetry of the underlying crystal structure (space group $P4_132$) was carried out using \textsc{basireps}~\cite{RODRIGUEZCARVAJAL199355} in the \textsc{fullprof} software package. Since the magnetic Bragg reflections were observed on top of the structural reflections as shown in the inset of Fig.~\ref{fig5}(a), the magnetic propagation vector $\vec{k}$ is equal to $(0,0,0)$. For the Wyckoff position $12d$ of magnetic Cu$^{2+}$ ions with a total of 12 spins in the unit cell as shown in Figs.~\ref{fig1}(b) and \ref{fig7}(c), the decomposition of the irreducible representations (IRs) can be described by
\begin{equation}
\Gamma = 1\Gamma_1^{(1)} + 2\Gamma_2^{(1)} + 3\Gamma_3^{(2)} + 4\Gamma_4^{(3)} + 5\Gamma_5^{(3)}\label{Ireps},
\end{equation}
where the basis vectors for $\Gamma_1$, $\Gamma_2$, $\Gamma_3$, $\Gamma_4$, and $\Gamma_5$ are given in Table~\ref{table4}.  $\Gamma_1$ and $\Gamma_2$ are one dimensional with one and two basis vector(s), respectively.  On the other hand, $\Gamma_3$ are of two dimensions with six basis vectors whereas $\Gamma_4$ and $\Gamma_5$ are of three dimensions with twelve and fifteen basis vectors, respectively.  We assume that there is only one order parameter for the magnetic transition in SrCuTe$_2$O$_6$ and hence based on the Landau theory~\cite{landau},  the magnetic structure of the low-temperature phase corresponds to a single IR. 

\begin{table*}
\begin{ruledtabular}
\caption{\label{table4}Magnetic irreducible representations and their basis vectors for Cu1$(x,y,z)$, Cu2$(-x+1/2,-y+1,z+1/2)$, Cu3$(-x+1,y-1/2,-z+1/2)$, Cu4$(x+1/2,-y+3/2,-z+1)$, Cu5$(z,x,y)$, Cu6$(z+1/2,-x+1/2,-y+1)$, Cu7$(-z+1/2,-x+1,y-1/2)$, Cu8$(-z+1,x+1/2,-y+3/2)$, Cu9$(y,z,x)$, Cu10$(-y+1,z+1/2,-x+1/2)$, Cu11$(y-1/2,-z+1/2,-x+1)$, Cu12$(-y+3/2,-z+1,x+1/2)$.}
\begin{tabular}{l c c c c c c c c c c c c c}

           IRs   & BV & Cu1  & Cu2 & Cu3 & Cu4 & Cu5 & Cu6 & Cu7 & Cu8 & Cu9 & Cu10 & Cu11 & Cu12\\
         \hline
             $\Gamma_1$ & $\psi_1$ & (10-1)\footnote{A parenthesis represents $(m_x m_y m_z)$.} & (-10-1) & (-101) & (101) & (-110) & (-1-10) & (1-10) & (110) & (0-11) & (0-1-1) & (01-1) & (011)\\
         \hline  
             $\Gamma_2$ & $\psi_1$ & (101) & (-101) & (-10-1) & (10-1) & (110) & (1-10) & (-1-10) & (-110) & (011) & (01-1) & (0-1-1) & (0-11)\\
				 & $\psi_2$ & (010) & (0-10) & (010) & (0-10) & (001) & (00-1) & (001) & (00-1) & (100) & (-100) & (100) & (-100)\\
	\hline
	     $\Gamma_3\footnote{The basis vectors for $\Gamma_3$ are complex with the first and second rows denoting the real and imaginary parts, respectively.}$ & $\psi_1$&(100)&(-100)&(-100)&(100)&(0-$\frac{1}{2}$0)&(0$\frac{1}{2}$0)&(0$\frac{1}{2}$0)&(0-$\frac{1}{2}$0)&(00-$\frac{1}{2}$)&(00$\frac{1}{2}$)&(00$\frac{1}{2}$)&(00-$\frac{1}{2}$)\\
				      &&(000)&(000)&(000)&(000)&(0-$\frac{\sqrt{3}}{2}$0)&(0$\frac{\sqrt{3}}{2}$0)&(0$\frac{\sqrt{3}}{2}$0)&(0-$\frac{\sqrt{3}}{2}$0)&(00$\frac{\sqrt{3}}{2}$)&(00-$\frac{\sqrt{3}}{2}$)&(00-$\frac{\sqrt{3}}{2}$)&(00$\frac{\sqrt{3}}{2}$)\\
				   & $\psi_2$&(010)&(0-10)&(010)&(0-10)&(00-$\frac{1}{2}$)&(00$\frac{1}{2}$)&(00-$\frac{1}{2}$)&(00$\frac{1}{2}$)&(-$\frac{1}{2}$00)&($\frac{1}{2}$00)&(-$\frac{1}{2}$00)&($\frac{1}{2}$00)\\
				      &&(000)&(000)&(000)&(000)&(00-$\frac{\sqrt{3}}{2}$)&(00$\frac{\sqrt{3}}{2}$)&(00-$\frac{\sqrt{3}}{2}$)&(00$\frac{\sqrt{3}}{2}$)&($\frac{\sqrt{3}}{2}$00)&(-$\frac{\sqrt{3}}{2}$00)&($\frac{\sqrt{3}}{2}$00)&(-$\frac{\sqrt{3}}{2}$00)\\
				   & $\psi_3$&(001)&(001)&(00-1)&(00-1)&(-$\frac{1}{2}$00)&(-$\frac{1}{2}$00)&($\frac{1}{2}$00)&($\frac{1}{2}$00)&(0-$\frac{1}{2}$0)&(0-$\frac{1}{2}$0)&(0$\frac{1}{2}$0)&(0$\frac{1}{2}$0)\\
				      &&(000)&(000)&(000)&(000)&(-$\frac{\sqrt{3}}{2}$00)&(-$\frac{\sqrt{3}}{2}$00)&($\frac{\sqrt{3}}{2}$00)&($\frac{\sqrt{3}}{2}$00)&(0$\frac{\sqrt{3}}{2}$0)&(0$\frac{\sqrt{3}}{2}$0)&(0-$\frac{\sqrt{3}}{2}$0)&(0-$\frac{\sqrt{3}}{2}$0)\\
				   & $\psi_4$&(00$\frac{1}{2}$)&(00$\frac{1}{2}$)&(00-$\frac{1}{2}$)&(00-$\frac{1}{2}$)&($\frac{1}{2}$00)&($\frac{1}{2}$00)&(-$\frac{1}{2}$00)&(-$\frac{1}{2}$00)&(0-10)&(0-10)&(010)&(010)\\
				      &&(00-$\frac{\sqrt{3}}{2}$)&(00-$\frac{\sqrt{3}}{2}$)&(00$\frac{\sqrt{3}}{2}$)&(00$\frac{\sqrt{3}}{2}$)&($\frac{\sqrt{3}}{2}$00)&($\frac{\sqrt{3}}{2}$00)&(-$\frac{\sqrt{3}}{2}$00)&(-$\frac{\sqrt{3}}{2}$00)&(000)&(000)&(000)&(000)\\
				   & $\psi_5$&(0$\frac{1}{2}$0)&(0-$\frac{1}{2}$0)&(0$\frac{1}{2}$0)&(0-$\frac{1}{2}$0)&(00$\frac{1}{2}$)&(00-$\frac{1}{2}$)&(00$\frac{1}{2}$)&(00-$\frac{1}{2}$)&(-100)&(100)&(-100)&(100)\\
				      &&(0-$\frac{\sqrt{3}}{2}$0)&(0$\frac{\sqrt{3}}{2}$0)&(0-$\frac{\sqrt{3}}{2}$0)&(0$\frac{\sqrt{3}}{2}$0)&(00$\frac{\sqrt{3}}{2}$)&(00-$\frac{\sqrt{3}}{2}$)&(00$\frac{\sqrt{3}}{2}$)&(00-$\frac{\sqrt{3}}{2}$)&(000)&(000)&(000)&(000)\\
				   & $\psi_6$&($\frac{1}{2}$00)&(-$\frac{1}{2}$00)&(-$\frac{1}{2}$00)&($\frac{1}{2}$00)&(0$\frac{1}{2}$0)&(0-$\frac{1}{2}$0)&(0-$\frac{1}{2}$0)&(0$\frac{1}{2}$0)&(00-1)&(001)&(001)&(00-1)\\
				      &&(-$\frac{\sqrt{3}}{2}$00)&($\frac{\sqrt{3}}{2}$00)&($\frac{\sqrt{3}}{2}$00)&(-$\frac{\sqrt{3}}{2}$00)&(0$\frac{\sqrt{3}}{2}$0)&(0-$\frac{\sqrt{3}}{2}$0)&(0-$\frac{\sqrt{3}}{2}$0)&(0$\frac{\sqrt{3}}{2}$0)&(000)&(000)&(000)&(000)\\

	\hline
	    $\Gamma_4$ & $\psi_1$     & (100) & (-100) & (100) & (-100)   & (000) & (000) & (000) & (000) & (0-10) & (010) & (0-10) & (010)\\
				& $\psi_2$     & (010) & (0-10) & (0-10) & (010)   & (000) & (000) & (000) & (000)  & (-100) & (-100) & (100) & (100) \\
				& $\psi_3$     & (001) & (001) & (001) & (001)      &(000) & (000) & (000) & (000)  & (00-1) & (00-1) & (00-1) & (00-1) \\   
				& $\psi_4$     & (000) & (000) & (000) & (000)      & (-110) & (110) & (1-10) & (-1-10) & (000) & (000) & (000) & (000) \\ 
				& $\psi_5$     & (00-1) & (001) & (0-1) & (001)     & (010) & (0-10) & (010) & (0-10)  & (000) & (000) & (000) & (000) \\
				& $\psi_6$     & (0-10) & (0-10) & (010) & (010)   & (001) & (00-1) & (00-1) & (001)  & (000) & (000) & (000) & (000) \\
				& $\psi_7$     & (-100) & (-100) & (-100) & (-100) & (100) & (100) & (100) & (100)  & (000) & (000) & (000) & (000)\\
				& $\psi_8$     & (000) & (000) & (000) & (000)      & (000) & (000) & (000) & (000) & (0-11) & (011) & (01-1) & (0-1-1) \\
				& $\psi_9$     & (000) & (000) & (000) & (000)      & (-100) & (100) & (-100) & (100) & (001) & (00-1) & (001) & (00-1) \\
				& $\psi_{10}$ & (000) & (000) & (000) & (000)     & (00-1) & (00-1) & (001) & (001) & (100) & (-100) & (-100) & (100) \\ 
				& $\psi_{11}$ & (000) & (000) & (000) & (000)      & (0-10) & (0-10) & (0-10) & (0-10) & (010) & (010) & (010) & (010) \\
				& $\psi_{12}$ & (10-1) & (101) & (-101) & (-10-1) & (000) & (000) & (000) & (000) & (000) & (000) & (000) & (000) \\		
	\hline
	   $\Gamma_5$  & $\psi_1$ & (100) & (-100) & (100) & (-100) & (000) & (000) & (000) & (000) & (010) & (0-10) & (010) & (0-10)\\
				& $\psi_2$ & (010) & (0-10) & (0-10) & (010) & (000) & (000) & (000) & (000) & (100) & (100) & (-100) & (-100) \\
				& $\psi_3$ & (001) & (001) & (001) & (001) & (000) & (000) & (000) & (000) & (001) & (001) & (001) & (001) \\
				& $\psi_4$ & (000) & (000) & (000) & (000) & (110) & (-110) & (-1-10) & (1-10) & (000) & (000) & (000) & (000) \\ 
				& $\psi_5$ & (000) & (000) & (000) & (000) & (001) & (001) & (001) & (001) & (000) & (000) & (000) & (000) \\
				& $\psi_6$ & (001) & (00-1) & (001) & (00-1) & (010) & (0-10) & (010) & (0-10) &(000) & (000) & (000) & (000) \\
				& $\psi_7$ & (010) & (010) & (0-10) & (0-10) & (001) & (00-1) & (00-1) & (001) & (000) & (000) & (000) & (000)\\
				& $\psi_8$ & (100) & (100) & (100) & (100) & (100) & (100) & (100) & (100) & (000) & (000) & (000) & (000) \\
				& $\psi_9$ & (000) & (000) & (000) & (000) & (000) & (000) & (000) & (000) & (011) & (0-11) & (0-1-1) & (01-1) \\
				& $\psi_{10}$ & (000) & (000) & (000) & (000) & (000) & (000) & (000) & (000) & (100) & (100) & (100) & (100) \\ 
				& $\psi_{11}$ & (000) & (000) & (000) & (000) & (100) & (-100) & (100) & (-100) & (001) & (00-1) & (001) & (00-1) \\
				& $\psi_{12}$ & (000) & (000) & (000) & (000) & (001) & (001) & (00-1) & (00-1) & (100) & (-100) & (-100) & (100) \\	
				& $\psi_{13}$ & (000) & (000) & (000) & (000) & (010) & (010) & (010) & (010) & (010) & (010) & (010) & (010) \\ 
				& $\psi_{14}$ & (101) & (10-1) & (-10-1) & (-101) & (000) & (000) & (000) & (000) & (000) & (000) & (000) & (000) \\
				& $\psi_{15}$ & (010) & (010) & (010) & (010) & (000) & (000) & (000) & (000) & (000) & (000) & (000) & (000) \\	
				
\end{tabular}
\end{ruledtabular}
\end{table*}

\end{document}